\documentclass[]{aastex631}
\usepackage{CJK}

\shorttitle{Rossby Number}
\shortauthors{Jao et al.}
\begin{document}
\begin{CJK*}{UTF8}{bsmi}
\title{Estimating the Convective Turnover Time}

\correspondingauthor{Wei-Chun Jao}
\email{wjao@gsu.edu}

\author[0000-0003-0193-2187]{Wei-Chun Jao (饒惟君)}
\affil{Department of Physics and Astronomy, Georgia State University, Atlanta, GA 30303, USA}

\author[0000-0001-9834-5792]{Andrew A. Couperus}
\affil{Department of Physics and Astronomy, Georgia State University, Atlanta, GA 30303, USA}

\author[0000-0002-1864-6120]{Eliot H. Vrijmoet}
\affil{Department of Physics and Astronomy, Georgia State University, Atlanta, GA 30303, USA}

\author[0000-0002-8389-8711]{Nicholas J Wright}
\affil{Astrophysics Group, Keele University, Keele, ST5 5BG, UK}

\author[0000-0002-9061-2865]{Todd J. Henry}
\affil{RECONS Institute, Chambersburg, PA 17201, USA}

\begin{abstract}

The introduction of the Rossby number (R$_0$), which incorporates the convective turnover time ($\tau$), in 1984 was a pioneering idea for understanding the correlation between stellar rotation and activity. The convective turnover time, which cannot be measured directly, is often inferred using existing $\tau$-mass or $\tau$-color relations, typically established based on an ensemble of different types of stars by assuming that $\tau$ is a function of mass. In this work, we use {\it Gaia} Early Data Release 3 to demonstrate that the masses used to establish one of the most cited $\tau$-mass relations are overestimated for G type dwarfs and significantly underestimated for late M dwarfs, offsets that affect studies using this $\tau$-mass relation to draw conclusions.  We discuss the challenges of creating such relations then and now. In the era of {\it Gaia} and other large datasets, stars used to establish these relations require characterization in a multi-dimensional space, rather than via the single-characteristic relations of the past. We propose that new multi-dimensional relations should be established based on updated theoretical models and all available stellar parameters for different interior structures from a set of carefully vetted single stars, so that the convective turnover time can be estimated more accurately.

\end{abstract}


%% Start the main body of the article. If no sections in the 
%% research note leave the \section call blank to make the title.

\section{Introduction}

The relation between stellar activity and rotation has been studied for over half a century, since \citet{Kraft1967} pioneered the work of showing that Ca II activity in early type stars is associated with their rotation. We now understand that this relation is closely related to stellar convection, differential rotation, magnetic field strength, and age \citep{Skumanich1972, Stix1976, Gilman1980, Noyes1984}. For example, in fast-rotating stars with a convective zone at the surface, the stellar dynamo generates magnetic fields that emerge above the photosphere. These twisted magnetic fields drive the heating of the atmosphere and can generate star spots or flares, which can then be detected photometrically or spectroscopically at different wavelengths \citep{Stepien1994, Wright2011, Newton2017, Jeffers2018}. In the past fifty years, the number of stars used to study this relation has increased dramatically, from less than one hundred stars to thousands, and the stellar types under consideration now stretch from early F type dwarfs to late M dwarfs. 

To understand the activity-rotation relation using an ensemble of stars with different masses and ages that have various rotation periods, surface activity levels, and convective zone depths, one often uses a dimensionless value, Rossby number (R$_0$). Commonly used in fluid dynamics, R$_0$ is defined as $U/L\Omega$ where $U$ is the fluid velocity, $L$ is a length over which the flow extends or exhibits variations, and $\Omega$ is the angular frequency. This ratio provides a rough estimate of the convective accelerations relative to the Coriolis force. In astronomy, this term is often given as $P_{rot}/\tau$ \citep{Noyes1984}, where P$_{rot}$ is the rotational period ($P_{rot}=1/\Omega$) and $\tau$ is the convective turnover time ($\tau=L/U$).  In practice, P$_{rot}$ is relatively easy to measure, whereas estimating an accurate $\tau$ is quite difficult, resulting in large uncertainties in derived Rossby numbers.

\citet{Gilman1980} stated that Rossby number is ``the most important single parameter determining what kind of differential rotation is to be expected in a stellar convection zone''. As a fluid element moves in the convection zone, a large ratio indicates that Coriolis forces have minimal time to act on an element and can be neglected. A small Rossby number indicates a larger impact from Coriolis forces, which can push flux rings from low altitudes to the direction of the poles \citep{Choudhuri1987}. Observationally, results indicate that activity levels measured using Ca II lines, H$\alpha$ emission, or x-rays saturate near a critical value of R$_{0c}\approx0.1 - 0.2$, and stellar activity decreases as R$_0$ increases \citep{Wright2011, Newton2017, Mittag2018}.

However, in the nearly forty years since \citet{Noyes1984} used R$_{0}$ to study the stellar activity-rotation relation, concerns have arisen about the effectiveness of using the Rossby number, e.g., as discussed in \citet{Stepien1994}, \citet{Reiners2014}, and \citet{Basri2021} (and references therein). Clearly, if one wants to use the Rossby number, the key challenge is to how to assign an appropriate value to the convective turnover time.  An in-depth discussion of whether or not to adopt the Rossby number is beyond the scope of this work, and is dependent on its specific application, but here we present the challenges of using the current established relations to estimate $\tau$, and provide suggestions for how to improve these relations in the future. 

%%%%%%%%%%%%%%%%%%%%%%%%%%%%%%%%%%
\section{Convective turnover time}

The convective turnover time, $\tau$, is derived from the mixing-length theory of stellar convection zones \citep{Prandtl1925, Bohm1958}. It is defined as $\Lambda$/$v_{conv}$, where $\Lambda$ is the mixing length at the base of the convection zone, and $v_{conv}$ is the convective velocity evaluated at $\Lambda$/2 above the base \citep{Gilliland1985}. The convective turnover time cannot be measured directly, so it is often inferred from stellar rotation and activity data. Because the depth of the convective zone depends on stellar mass, one often determines $\tau$ empirically by dividing an ensemble of stars into different bins of color as a proxy for mass. Scaling the rotation periods with a color-dependent or mass-dependent $\tau$ can minimize the scatter in the rotation-activity relation, and eventually produce a simple broken power-law fit relating activity to Rossby number \citep{Kiraga2007, Wright2011, Wright2018, Mittag2018}. Typically, the mean color or mass in each bin is used to establish an empirical relation of $\tau = f(color)$, where color is often $B-V$ or $V-K_s$, or $\tau = f(M)$, where $M$ is stellar mass, if masses can be estimated. Such an empirical relation can then be easily applied to different samples of stars to obtain their convective turnover times if their colors or masses are available. Using $B-V$=0.65 and $V-K_s$=1.51 for the Sun as an example, we calculate the convective turnover times from three empirical relations, and find results that span $\tau$ $\sim$ 10.4--49.5 days, as shown in Table~\ref{tbl:Sun}. The range in these values mimics a larger range in values between 5 and 45 derived using helioseismology, the TGEC evolutionary model for the Sun, and a Standard Solar Model, also given in Table~\ref{tbl:Sun}. Thus, even the $\tau$ value for the Sun is poorly constrained between values derived from various methods, implying that our knowledge of convective turnover times for other stars is limited.

\begin{deluxetable}{lcccc}
\tablecaption{Convective turnover times for the Sun\label{tbl:Sun}}
\tablehead{
\colhead{} &
\colhead{Type} &
\colhead{$\tau$-color} &
\colhead{$\tau$} &
\colhead{Ref.}\\
\colhead{} &
\colhead{} &
\colhead{relation} &
\colhead{$(days)$} &
\colhead{}
}
\startdata
\hline
Wright+ (2018)  & empirical & $\tau$-($V-K_s$) & 10.4 &(1) \\
Mittag+ (2018)  & empirical & $\tau$-($B-V$) & 35.4 &(2)\\
Corsaro+ (2021) & empirical & $\tau$-($B-V$) & 49.5 &(3)\\
helioseismology & & & 4.9$+$ & (7)\\
TGEC & theoretical & $\tau$-($B-V$) & 16.5 & (8) \\
Standard Solar Models & theoretical & & 30-45 & (3, 4, 5, 6)
\enddata
\tablecomments{The convective turnover time from helioseismology is calculated based on R$_{0}\approx$5 just below the photosphere of the Sun \citep{Greer2016} and the rotation period of 24.5 days at the equator. \citet{Greer2016} showed the Sun's Rossby number quickly drops to 0.4 at a depth of 10 million meters or 0.01R$_{\odot}$. The Toulouse-Geneva stellar evolution code (TGEC) also provides a $\tau -(B-V)$ relation to calculate the Rossby number, even though this relation is established from models.}
\tablerefs{(1)=\citet{Wright2018}, (2)=\citet{Mittag2018}, (3)=\citet{Corsaro2021}, (4)=\citet{Brun1999}, (5)=\citet{Bonanno2002}, (6)=\citet{Landin2010}, (7)=\citet{Greer2016}. (8)=\citet{Castro2014}}
\end{deluxetable}

%%%%%%%%%%%%%%%%%%%%%%%%%%%%%%%%%%
\section{Example of an ensemble of stars used to determine a $\tau$-mass relation}

\begin{figure}[h!]
\begin{center}
\includegraphics[scale=0.7,angle=0]{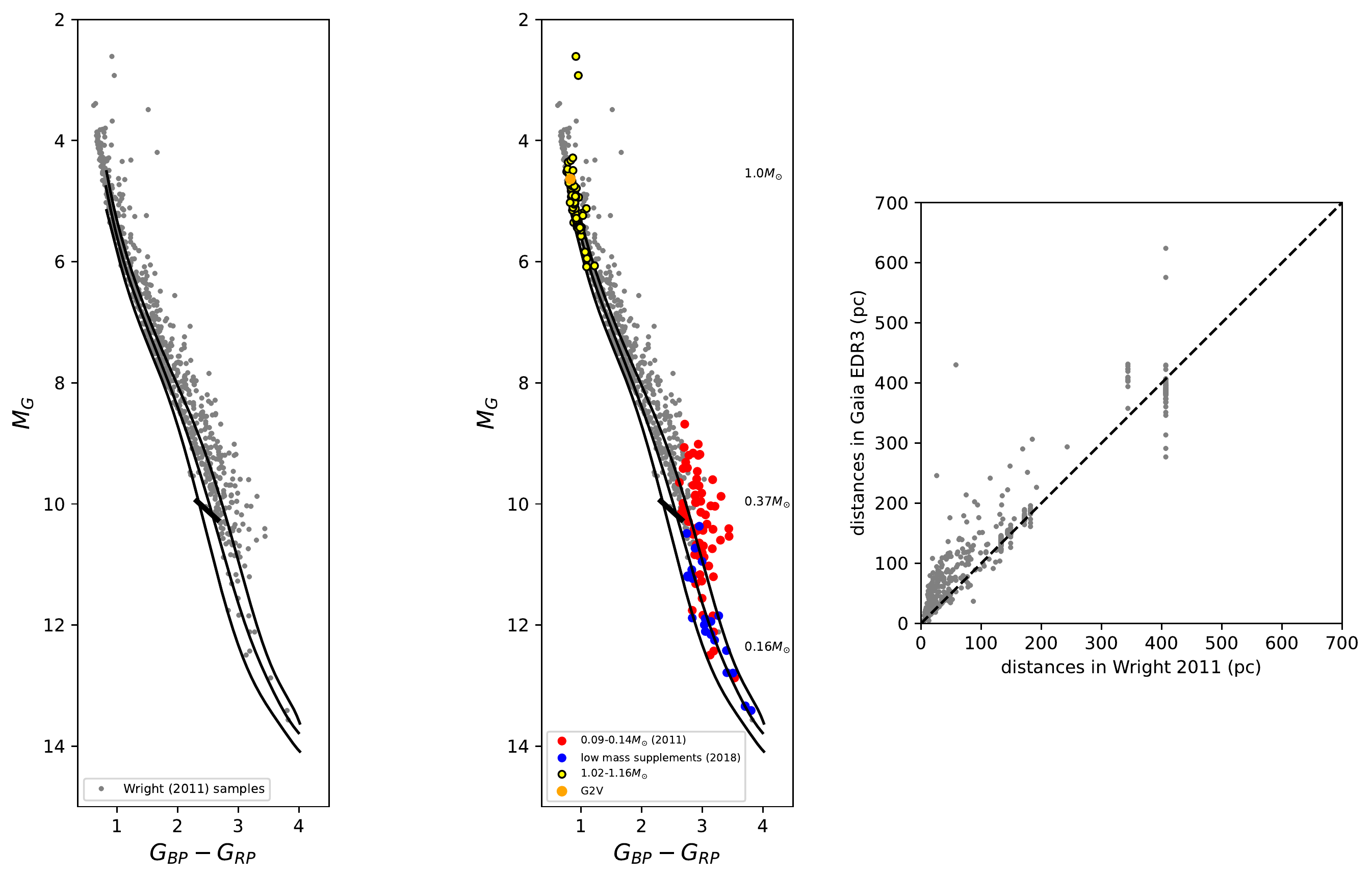}
\caption{(Left) Stars in \citet{Wright2011} used to generate the $\tau$-mass relation are shown on an observational HR diagram using {\it Gaia} photometry and parallaxes. The central black line represents a fit to the distribution of stars on the main sequence in {\it Gaia} EDR3 within 100 parsecs, and the upper and lower black lines encompass 90\% of the population at a given color \citep{Jao2022}. The short, thick black line represents the location of the main sequence gap \citep{Jao2020}, corresponding to the transition between partially and fully convective M dwarfs. (Center) Yellow dots are stars in the mass bin of 1.02 to 1.16 M$_{\odot}$ in \citet{Wright2011} and an orange dot represents a G2V dwarf as a reference. Red dots are stars with masses included in the lowest mass bin with masses of 0.09 to 0.14M$_{\odot}$ in \citet{Wright2011}, and blue dots are supplemental low mass stars presented in \citet{Wright2018}. Three masses of 1.0, 0.37, and 0.16M$_{\odot}$ are labeled at their corresponding spectral types of G2V, M3V, and M5V based on the stellar parameters table in \citet{Pecaut2013}. (Right) 
A comparison between distances from {\it Gaia} EDR3 and distances used in \citet{Wright2011} to derive their fundamental parameters; most distances were underestimated. \label{fig:WrightHRD}}
\end{center}
\end{figure}

The Hertzsprung-Russell Diagram (HRD) in Figure~\ref{fig:WrightHRD} shows the sample of stars with x-ray measurements given in \citet[hereafter W11]{Wright2011}, where a $\tau$-mass relation was derived that could be applied to stars with masses less than 1.36$M_\odot$. Among the 824 stars in W11, we match 809 (97\%) to stars in {\it Gaia} EDR3 \citep{EDR3} results. Most stars are above the gap in the main-sequence marked as a thick line in Figure~\ref{fig:WrightHRD}, corresponding to the transition to fully convective low-mass stars \citep{Jao2018}.  Thus, most of the points in the plot represent partially convective stars with masses greater than 0.32-0.36$M_{\odot}$ \citep{Saders2012, Baraffe2018}. A few additional fully convective stars with masses less than $\sim$0.35$M_{\odot}$ below the gap have been added by recent efforts \citep[hereafter W18]{Wright2016, Wright2018}, and the $\tau$-mass relation was rederived, but there are still many more stars above the gap than below. The lack of low-mass, fully convective stars in this sample is mainly because stars below the gap are very faint, particularly in x-rays, so targeted observations are required. Even if x-ray observations are secured, some low mass M dwarfs may have rotation periods longer than 100 days, requiring considerable observing time investments to determine their periods. Finally, as can be seen in Figure~\ref{fig:WrightHRD} and summarized in W11, most of these stars with x-ray detections are elevated above the main sequence, confirming that they are young stars selected from nearby young moving clusters.

In W11, the entire sample was divided into ten $V-K_s$ or mass bins corresponding to approximate masses of 0.09 to 1.36$M_{\odot}$ with varying mass bin sizes of 0.04 to 0.21$M_{\odot}$. Stars falling in the last bin with masses of 0.09--0.14$M_{\odot}$ are shown as red dots in Figure~\ref{fig:WrightHRD}. For reference, three approximate mass guidelines for spectral types of G2V, M3V, and M5V as listed in \citet{Pecaut2013} are labeled in the middle panel of Figure~\ref{fig:WrightHRD}\footnote{See the table at https://www.pas.rochester.edu/$\sim$emamajek/EEM\_dwarf\_UBVIJHK\_colors\_Teff.txt, which has been updated by E.~Mamajek since 2013.}. We can see that masses in the last bin outlined in W11 are typically underestimated by a factor of two; in fact, 75 of the 79 stars in this mass bin are plotted here, but none has a mass less than 0.14$M_{\odot}$ according to \citet{Pecaut2013}. Another group of stars with masses in the 1.02 to 1.16 $M_{\odot}$ bin is shown as yellow dots in Figure~\ref{fig:WrightHRD}. It appears that these stars are likely less massive than 1$M_{\odot}$ rather than 1.02--1.16$M_{\odot}$ because most are redder and less luminous than the Sun, shown with the large yellow point on the HRD.

To establish a robust $\tau$-mass relation requires reliable masses. Determining masses was a challenging task in 2011 because few cluster and field stars, including binaries for which masses could be determined, had accurate trigonometric parallaxes.  So, the 824 stars with x-ray measurements in W11 typically had only estimated distances, and consequently, mass estimates. Even for stars in the same cluster with parallax measurements, W11 used a fixed distance for all members in the cluster. For field stars without distances, multiple steps were applied to estimate distances by converting $V-K_s$ to effective temperature using lookup tables and isochrones, and assuming their ages to be 1 Gyr. In the right plot of Figure~\ref{fig:WrightHRD}, we give a comparison between the distances used in W11 and those from {\it Gaia} EDR3, showing that the majority of distances were underestimated, thus degrading the reliability of any derived $\tau$-mass relation because of poor mass estimates.
Even with later supplements of a few additional late M dwarfs in W18 to re-establish the $\tau$-mass relation, the bulk of the data are still from the sample used in W11. Finally, we note that W18 reported only 35 stars in the last mass bin, down from 79 stars in W11, but Figure~\ref{fig:WrightHRD} shows only a few of those 35 stars actually belong in this mass bin. 

During the past decade, studies of activity-rotation relations have been extended to include late, fully-convective M dwarfs because of the {\it Kepler/K2} and {\it TESS} missions, which have permitted evaluation of flaring activity and rotation periods for thousands of nearby M dwarfs \citep{Davenport2016, Gunther2020, Raetz2020}. Existing $\tau$-mass and/or $\tau$-color relations have been used to obtain the Rossby numbers for many of these stars, but these relations should be revised using current {\it Gaia} data to secure more reliable estimates of the convective turnover time, particularly in the fully convective regime.

\section{Challenges of establishing a convective turnover time relation}

There are several challenges to establishing a convective turnover time relation that might be overcome with the advent of new data and techniques, particularly for the M dwarfs that are the focus of the following discussion.  These include accurate colors and luminosities using {\it Gaia} data, ages, metallicities, magnetic fields, sample vetting, and considerations of the partially/fully convective boundary.

{\it Improved Colors and Luminosities} --- Before the \textit{Gaia} mission, the vast majority of M dwarfs did not have parallaxes due to their intrinsic faintness, and consequently only a limited number of nearby stars could be plotted on the HRD. The best way to separate those stars into mass bins to establish $\tau$ relations was by using their colors, spectral types, or effective temperatures, any of which might be used to estimate masses. Typically, the relations were derived using stars grouped using only one of these parameters.  The high precision parallaxes and photometry from {\it Gaia} have changed how we can group stars that are within {\it Gaia's} observing limits,
creating a rich two-dimensional observational HRD with accurate colors and absolute magnitudes that can be explored with respect to activity levels, ages, and complex stellar interiors. As an example, consider slicing the HRD vertically to include stars falling between $BP-RP$ colors of 2.3 and 2.8.  This region includes the widest part of the main-sequence, both partially convective and fully convective stars, and young stars above the main sequence, as shown in Figure~\ref{fig:WrightHRD}. Previous studies determining empirical $\tau$ relations grouped this heterogeneous mix of stars together using binned colors, assuming that all stars had the same stellar properties, including masses, whereas the width of the main sequence means that these stars are far from identical. Slicing the HRD horizontally for $M_G$ = 9.5--10.5 again reveals that a wide range of stars is collected in one sweep, from cool subdwarfs to pre-main sequence stars that have different metallicities, radii, and convective depths. Alas, simply using one parameter to estimate masses for a group of stars is not enough. Another option is to use established mean mass-luminosity relations (MLR) \citep{Torres2010, Benedict2016, Mann2019} for sets of individual stars, but these relations all include stars of various metallicities, activity levels, and ages, and are not applicable to the young stars that are often used to study activity-rotation relations. Thus, it is difficult to estimate accurate stellar masses for these stars by simply applying available MLRs, and these inaccuracies propagate when determining the convective turnover time.  However, the recently released Gaia-DR3 non-single star catalog and low resolution spectra of millions of stars \citep{NSS, BPRPspectra} may provide a path to establish a metallicity dependent MLR, so that a proper MLR can be applied to those single active stars.

{\it Ages} --- \cite{GaiaHRD2018} demonstrated a composite HRD of open clusters and globular clusters with different ages and metallicities. That means two stars with the same colors may have different ages and metallicities, so the activity level should be different. Therefore, knowing age or metallicity of stars used to establish these relations  would be essential. An accurate estimate of the ages of coeval stars in clusters or associations is more reliable compared to the results for field stars, but the age of field stars can be estimated using gyrochonology \citep{Barnes2007}. However, despite the increasing number of low mass stars with identified rotation periods \citep{Popinchalk2021}, as well as recent efforts to understand the rotation-age relation for M dwarfs \citep{Rebull2018}, reliable gyrochonology relations to estimate M dwarf ages are still largely missing \citep{Angus2019}.

{\it Metallicities} --- As for obtaining metallicities for these active stars, although these stars could have metallicities in the literature, they are not measured in a uniform way. The challenge would be obtaining their spectra and re-measuring metallicities in a more consistent manner, but the release of $Gaia$ spectra and spectroscopic parameters \citep{Fouesneau2022} could provide a key step toward this goal.

{\it Magnetic Fields} --- To estimate a single active star's mass is challenging, so often a mass is estimated using isochrones from theoretical models, like the method used in \citet{Wright2011}.  However, all stars used to establish the $\tau$ relations are active, and studies show these active stars could have inflated radii and lowered temperatures \citep{Somers2017, Parsons2018, Jackson2019}, which then changes their luminosities and colors compared to inactive stars with the same masses. Recently, \citet{Simon2019} found that isochrones of pre-main sequence stars that don't include internal magnetic fields underestimate the dynamical masses for stars between 0.4 and 1.4$M_{\odot}$ by 30\%. Using isochrones incorporating magnetic fields may improve the average difference between a dynamical mass and the estimated isochrone track mass significantly, down to about 0.01$M_{\odot}$. Lately, \cite{Flores2022} reported 40 very young T Tauri stars masses between 0.3 and 1.3$M_{\odot}$ using both optical and infrared spectra after considering magnetic fields. They found that the spectroscopically-derived mass averagely differ from the dynamical masses by about 12 and 8\% in optical and infrared band, respectively. However, for stars less than 0.5M$_{\odot}$, they found the masses derived using infrared spectra are overpredicted by 31\%, and the masses derived using optical spectra astonishingly are overpredicted by 94\%. Consequently, understanding magnetic fields for these active stars with various ages poses another challenge to estimate their masses \citep{Feiden2016}.

{\it Sample Vetting} --- The samples used to generate $\tau$-color or $\tau$-mass relations in the past often contained unresolved binaries that affect the color and mass values used to generate the relations.  Crosschecking sample stars with Washington Double Star Catalog \citep{WDS} entries for companions with separations less than 2$^{\prime\prime}$, we find that 14\% of stars in W11 and 18\% of stars in \citet{Mittag2018} are potential close multiples. In these cases, the combined photometry or spectra will affect color and mass estimates and confuse the sources of rotation periods, thereby corrupting derived $\tau$ relations.  Given the multiplicity rates of 47\% for FGK dwarfs \citep{Raghavan2010} and 27\% for M dwarfs \citep{Winters2019}, vetting close binaries is a necessary step to yield better relations.

{\it Considerations of the Partially/Fully Convective Boundary} --- Finally, the discovery of a gap in the distribution of stars on the main sequence \citep{Jao2018} has led to new insight into a class of slowly pulsating M dwarfs at the transition region between partially and fully convective stars. Some M dwarfs could have up to three layers of interior structure, including a convective zone at the surface, another convective zone at the core, and a radiative layer in between \citep{Saders2012, Baraffe2018}. Because of $^3$He fusion instabilities, the two convective layers could merge and the radiative zone would disappear at times, causing the luminosity to drop and the radius to decrease. These stars oscillate between one and three layers of interior structure, causing their convective turnover times to change significantly, even when positioned on the HRD main sequence. Such stars pose an additional challenge to the assumption that convective turnover times are a straightforward function of mass or color.

%%%%%%%%%
\section{Conclusion}

Understanding stellar activity-rotation relations is an important topic because it helps us to understand not only stellar dynamos, but how stellar activity affects exoplanets' formation, atmospheres, and habitability. While it is debatable whether it is advisable or not to use the Rossby number when studying stellar activity, if one decides to use it, a reliable convective turnover time, $\tau$, must be determined. Here we have demonstrated that (1) a widely-used $\tau$-mass relation has masses for late M dwarfs typically underestimated by at least a factor of two, (2) earlier efforts have included unresolved binaries mixed in their samples that affect derived quantities, (3) color or mass bins typically include very different types of stars, and (4) some populations on the main sequence have complex, and dynamic interior structures. All of these issues complicate our understanding of convective turnover times.

\begin{figure}
\begin{interactive}{js}{3D.zip}
\includegraphics[scale=0.7,angle=0]{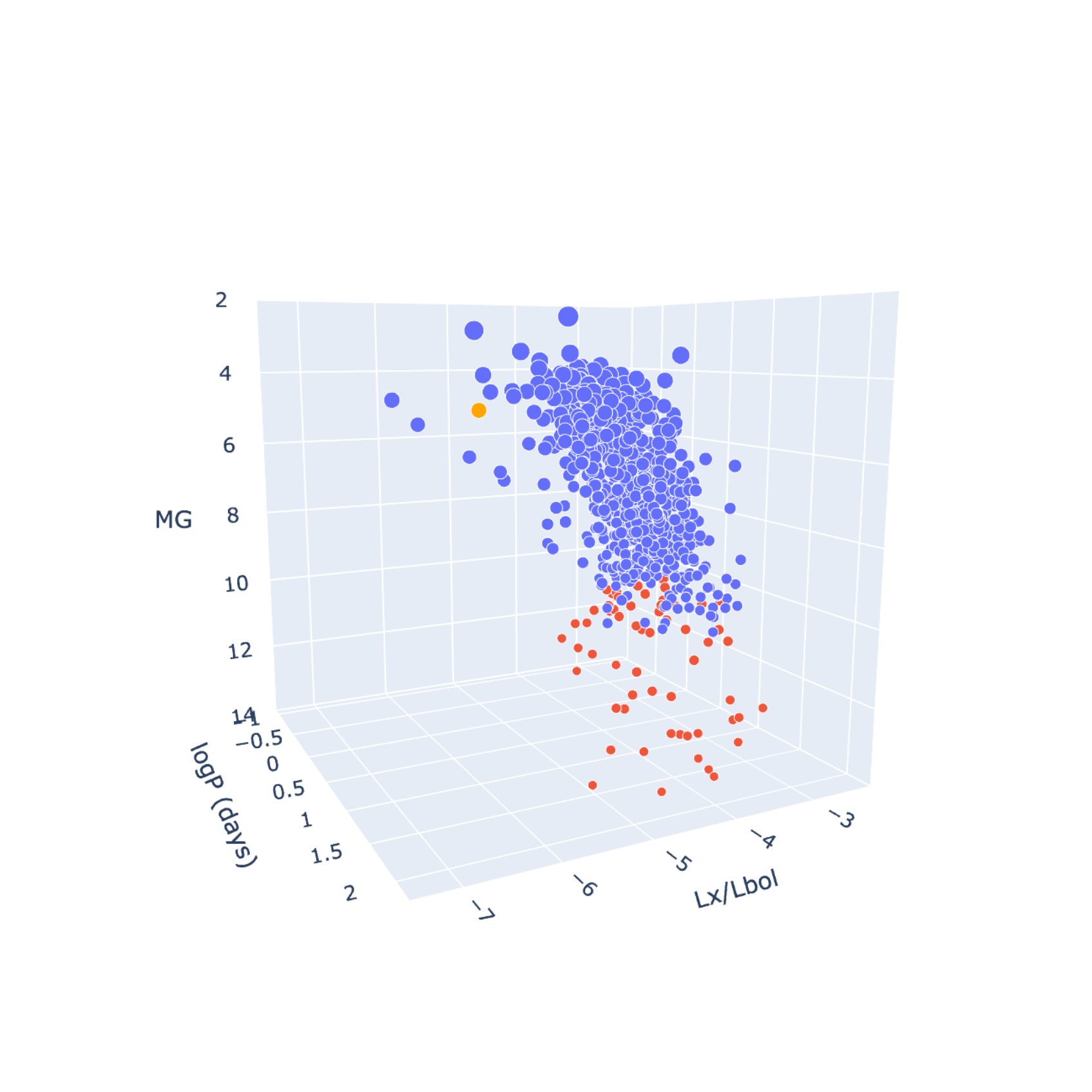}
\end{interactive}
\caption{A three dimensional interactive plot for data used in W11 and W18. The size of each point is for illustration only and does not show stars' true sizes relative to a G2V dwarf, although the large-to-small trend is correct. Blue dots are stars above the gap, seen as a thick black line in Figure~\ref{fig:WrightHRD}, and red dots are stars below the gap. An orange circle represents the Sun with $\log P=$1.4 and $L_{X}/L_{bol}=-$6.24 \citep{Wright2011}. This figure can be rotated and zoomed in and out, and the control panel of navigating this figure is available in the upper right corner of this figure.  An interactive version of this Figure is available in the online journal or \href{https://www.chara.gsu.edu/~jao/3D.html}{here}.}
\label{fig:3DHRD}
\end{figure}

Figure~\ref{fig:3DHRD} is a 3D interactive plot, illustrating rotational periods and x-ray luminosities presented in W11 and W18, plotted against $M_G$ values using {\it Gaia} EDR3 photometry and parallaxes. Blue points represent stars above the gap, and red points represent stars below it.  The size of each dot roughly represents the size of a star relative to a G2V dwarf (represented by the Sun with a yellow point) by scaling $M_G$.  These sizes are for illustration only, and the sizes for young stars in their sample are underestimated here, although this is at least a representation of a fourth parameter, stellar radius, that governs the distribution of points on this graph. A fifth parameter, not shown here, is the age or metallicity of each star, and it may be presented as a grid of multiple 3D plots like this graph. Such a complicated graph has been traditionally simplified by grouping all these stars into different mass or color bins, and then the optimal convective turnover time in each bin has been determined by minimizing the scatter on this graph.

In this paper, we have outlined some of the challenges to developing a reliable relation for convective turnover times.  Now that we are in the era of astronomical big data, we have accurate astrometry, stellar multiplicity information, broad- and narrow-band photometry from the ultraviolet to infrared, and spectra available for thousands to millions of stars, all augmented with better stellar evolutionary models. Many of the young cluster stars used to establish $\tau$ relations are also well-studied stars in the literature.  We propose that by utilizing the suite of available measurements, and a deep dive into the literature, it will be possible to overcome a few of these challenges, and as a result, an updated understanding of convective turnover times and Rossby numbers may be achieved.

\software{Matplotlib
  \citep{Hunter2007},
  NumPy \citep{vanderWalt2011}, Plotly \citep{Plotly2015}, and SciPy \citep{Virtanen2020}} 

\begin{acknowledgements}

We would like to thank Eric Mamajek, Jane Pratt, Gibor Basri and Petrus Martens for helpful discussions. This work was supported by the NASA Astrophysics Data Analysis Program (ADAP) under grant 20-ADAP20-0288. This research has made use of the SIMBAD database, operated at CDS, Strasbourg, France. This work has made use of data from the European Space Agency (ESA) mission {\it Gaia} (\url{https://www.cosmos.esa.int/gaia}), processed by the {\it Gaia} Data Processing and Analysis Consortium (DPAC, \url{https://www.cosmos.esa.int/web/gaia/dpac/consortium}).  Funding for the DPAC has been provided by national institutions, in particular the institutions participating in the {\it Gaia} Multilateral Agreement. This research has made use of NASA's Astrophysics Data System.

\end{acknowledgements}

\end{CJK*}
\end{document}